\documentclass[prb,aps,noshowpacs,preprintnumbers,amsmath,amssymb,twocolumn,groupedaddress,floatfix]{revtex4-2}
\usepackage{graphicx}
\usepackage{hyperref}
\usepackage{xcolor}
\usepackage{floatrow}
\usepackage[caption=false]{subfig}
\usepackage[switch]{lineno}

\begin{document}
\title{Flux-driven delocalization transition in disordered topological insulator nanowires}
\author{Shimon Arie Haver}
\thanks{These authors contributed equally: Shimon Arie Haver and Emuna Rimon.}
\affiliation{Department of Physics, Ben-Gurion University of the Negev, Be\textquoteright er-Sheva
84105, Israel}
\author{Emuna Rimon}
\thanks{These authors contributed equally: Shimon Arie Haver and Emuna Rimon.}
\affiliation{Department of Physics, Ben-Gurion University of the Negev, Be\textquoteright er-Sheva
84105, Israel}
\author{Eytan Grosfeld}
\email{grosfeld@bgu.ac.il}
\affiliation{Department of Physics, Ben-Gurion University of the Negev, Be\textquoteright er-Sheva
84105, Israel}
\date{\today}
\begin{abstract}
Topological insulator nanowires provide a tunable platform for studying the interplay between disorder, quantum interference, and symmetry-protected transport. Here we investigate quantum transport in disordered topological insulator nanowires threaded by an axial magnetic flux. 
By computing the conductance as a function of wire length, magnetic flux, chemical potential, and disorder strength, we extract the localization length to characterize the flux-driven delocalization transition near half-integer flux quanta. 
We find that the localization length diverges with a robust critical exponent \(\nu=2\), independent of the chemical potential and disorder strength considered here. 
This exponent differs from that of the integer quantum Hall transition, pointing to distinct scaling behavior. 
Near integer flux quanta, we further find that the conductance evolves from a weak-localization dip at low chemical potential to a weak anti-localization peak at higher chemical potential, which splits and is eventually suppressed as the system crosses over to the strongly localized regime.
\end{abstract}
\maketitle
\section{Introduction} 

Quantum transport in topological insulator nanowires (TINWs) offers a powerful platform to explore the interplay between topology, dimensionality, and disorder. Their low-energy physics is governed by a single-node massless Dirac equation constrained to the wire surface. When subjected to an axial magnetic field, TINWs display rich flux-dependent phenomena arising from the unique structure of their surface-state spectrum \cite{ostrovsky2010interaction,Bardarson2013}. Recent progress in fabricating bulk-insulating nanowires \cite{rossler2023top} has further enhanced the feasibility of probing these effects experimentally.

A hallmark of this system is the appearance of weak anti-localization (WAL) peaks in the conductance at integer flux quanta, a consequence of spin-momentum locking and quantum interference \cite{hikami1980spin}. At half-integer flux quanta, the surface states host a topologically protected helical mode that remains immune to non-magnetic backscattering due to time-reversal symmetry (TRS), leading to a distinctive conductance peak. Together, these mechanisms reshape the periodicity of Aharonov–Bohm oscillations \cite{bardarson2010aharonov,zhang2010anomalous,Bardarson2013,cho2015aharonov,Jauregui2016,munning2021quantum}. 

The axial flux thus provides a natural tuning knob that modulates transport and drives crossovers between localized and delocalized regimes \cite{cho2015aharonov}. The key quantity governing this behavior is the localization length, which controls the decay of conductance with wire length. In particular, the divergence of the localization length near the protected helical mode provides a way to characterize the corresponding delocalization transition.

\begin{figure}[b]
\begin{centering}
\includegraphics[width=0.8\linewidth]{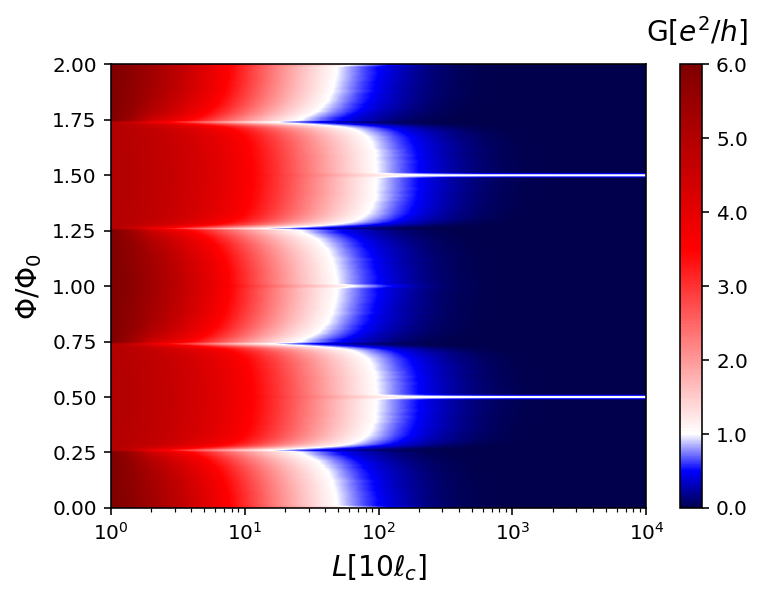}
\par\end{centering}
\caption{{\bfseries Conductance versus wire length and axial flux.}
\(G\) as a function of wire length \(L\) and dimensionless axial flux
\(\eta=\Phi/\Phi_0\), for \(\mu=2.751\,\hbar\Omega\) and \(W=0.05\,\hbar\Omega\),
averaged over \(1000\) disorder realizations.}
\label{fig:conductance-L-eta}
\end{figure}

In this work, we investigate the dependence of the localization length in TINWs on axial flux, chemical potential, and disorder strength. Near half flux quantum, where a perfectly transmitting helical mode emerges, we find a flux-driven delocalization transition characterized by a robust critical exponent. Near integer flux quantum, we analyze the weak-localization and weak anti-localization features and show that their behavior depends on the
number of participating modes. In particular, the conductance evolves from a
weak-localization dip at low chemical potential to a weak anti-localization
peak at higher chemical potential. We further show that, with increasing wire
length, the anti-localization peak splits and is eventually suppressed as the
system crosses over to the strongly localized regime.

These results demonstrate that axial flux in disordered TINWs provides more
than a periodic modulation of the conductance. It enables direct access to
the localization properties of the protected helical mode and to the
evolution of interference effects near integer flux quanta. The combination
of a flux-driven divergence of the localization length and a mode-dependent
weak-localization to weak anti-localization crossover establishes TINWs as a
controlled setting for studying how topology, symmetry, and disorder shape
quantum transport in a quasi-one-dimensional geometry.

\begin{figure*}[t]
\begin{centering}
\includegraphics[width=0.8\linewidth]{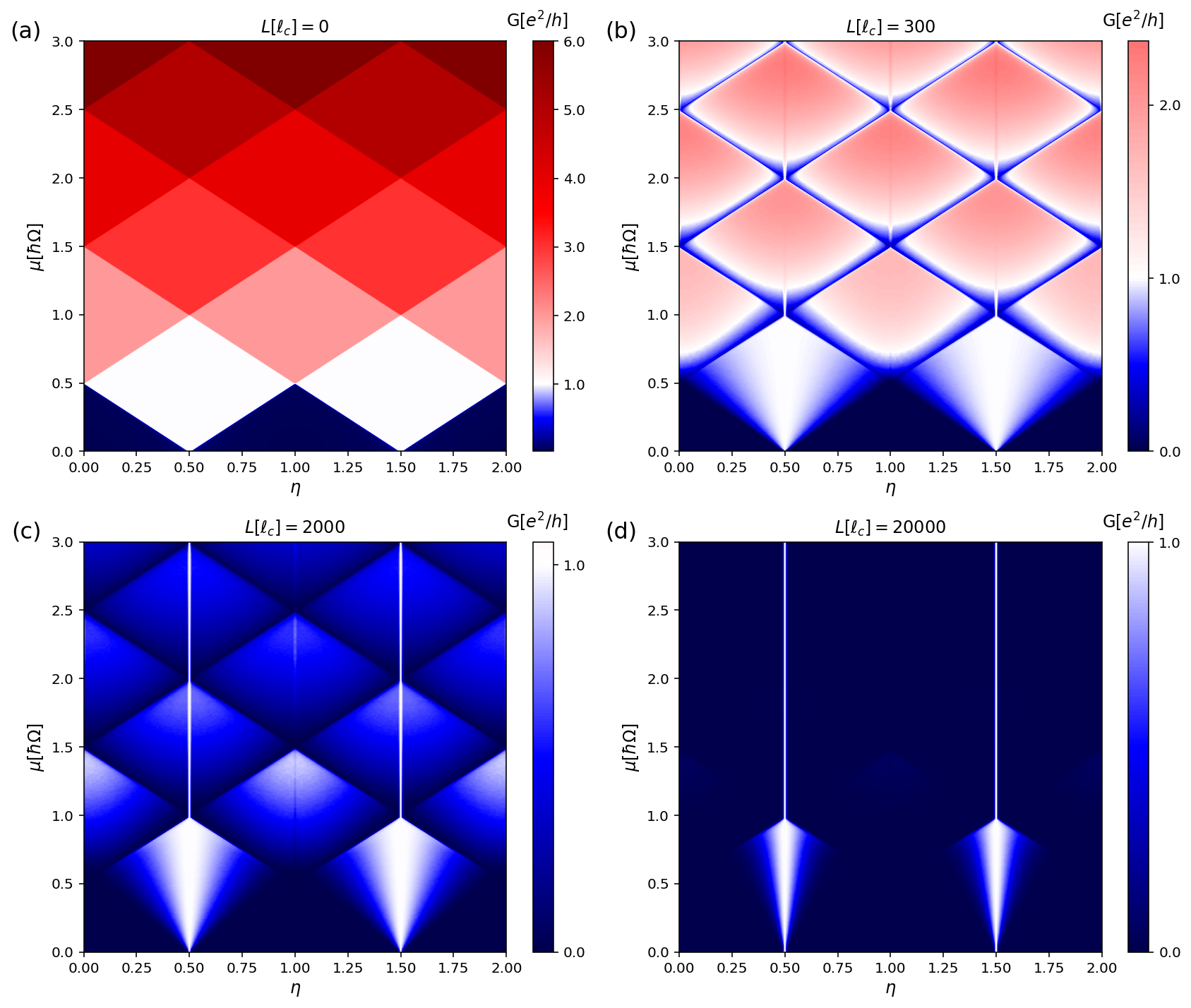}
\par\end{centering}
\caption{{\bfseries Conductance as a function of chemical potential and axial flux.}
\(G\) is shown versus chemical potential \(\mu\) and dimensionless axial flux
\(\eta=\Phi/\Phi_0\), for wire lengths
\(L/\ell_c=0\) (a), \(300\) (b), \(2000\) (c), and \(20000\) (d).
With increasing \(L\), the clean-limit diamond pattern is suppressed, while
a conductance peak remains near \(\eta=1/2\). The remaining parameters are
the same as in Fig.~\ref{fig:conductance-L-eta}.}
\label{fig:conductance-mu-eta}
\end{figure*}
\section{Results}

The TINW surface state is governed by a two-dimensional Dirac Hamiltonian with longitudinal momentum $k$ along the wire and orbital angular momentum $\ell$ around its circumference. The wire is threaded by a dimensionless magnetic flux $\eta=\Phi/\Phi_0$ where $\Phi_0$ is the flux quantum, and is filled up to chemical potential $\mu$. We measure energies in units of $\hbar\Omega$ where $\Omega=v_{\text{F}}/R$ with $R$ the wire radius and $v_{\text{F}}$ its Fermi velocity~\cite{haver2023electromagnetic}

The TINW contains randomly distributed impurities. Each impurity is characterized by a random scattering strength drawn independently from a uniform distribution $\mathcal{U}([-W,W])$. For short-range impurities, the scattering is taken to couple all orbital angular momentum channels uniformly. The average longitudinal distance between impurities is denoted by $\ell_c$. We compute the conductance $G$ using a scattering matrix approach, described in Methods, as a function of the wire length $L$.

\subsection{Unveiling the Localization Landscape} 

We first study the dependence of the conductance on the axial magnetic flux
and on the length of the wire. In the clean limit, changing the flux shifts
the surface-state spectrum and changes the number of propagating angular
momentum channels at the Fermi level. This produces the characteristic
oscillations of the conductance as a function of flux. Once disorder is
introduced, these oscillations are gradually suppressed with increasing wire
length, as shown in Fig.~\ref{fig:conductance-L-eta}. For sufficiently long
wires, most values of the magnetic flux lead to localization. The exception
occurs near half-integer flux quantum, where a perfectly conducting helical
channel remains visible even at the longest lengths considered.

This robust conducting channel appears when the number of propagating modes
is odd, and is protected by time-reversal symmetry
\cite{Bardarson2013,ando2002presence}. In addition, conductance peaks are
observed near integer flux quanta at intermediate wire lengths. These peaks
are associated with weak anti-localization in the time-reversal-invariant
points, but they are eventually suppressed as the system crosses over to the
strongly localized regime.

The clean-limit oscillations observed in Fig.~\ref{fig:conductance-L-eta}  and Fig.~\ref{fig:conductance-mu-eta}a can be understood from the number of conducting
channels at fixed chemical potential. For a given dimensionless chemical
potential \(\tilde{\mu}=\mu/(\hbar\Omega)\), the propagating angular momentum channels lie in the range $\ell_\text{min},\ldots,\ell_\text{max}$, with
\[
\ell_{\text{min}}
=
\left\lceil -\tilde{\mu}-\frac{1}{2}+\eta \right\rceil,
\qquad
\ell_{\text{max}}
=
\left\lfloor \tilde{\mu}-\frac{1}{2}+\eta \right\rfloor,
\]
so that the number of channels $n$ is
\[
n=\ell_{\text{max}}-\ell_{\text{min}}+1 .
\]
As the flux is varied at fixed \(\mu\), this number alternates between two integer values
\(n_0\) and \(n_0+1\), giving the clean conductance oscillations.

The dependence on chemical potential is shown in
Fig.~\ref{fig:conductance-mu-eta}. In the clean limit, the conductance forms
a diamond pattern in the \((\mu,\eta)\) plane, reflecting the opening and
closing of propagating angular momentum channels. At larger chemical
potential, more channels are available and the ballistic conductance is
larger. However, these higher-channel regimes are also more strongly affected
by disorder, and the distinction between neighboring diamonds is rapidly
washed out even at relatively short lengths.

The remaining chemical-potential dependence is strongest near the edges of
the diamonds, where the density of states is enhanced. In contrast, the
regime with a single conducting mode, which occurs at low chemical potential,
shows the slowest decay with length. As the wire becomes longer, the
ballistic diamond pattern is replaced by diffusive behavior with conductance peaks near integer and
half-integer flux quantum. At the longest lengths, strong localization sets in, and only the peaks associated
with the helical channel near half-integer flux remain.

\subsection{The weak anti-localization peaks}

\begin{figure}[t]
\begin{centering}
\includegraphics[width=\linewidth]{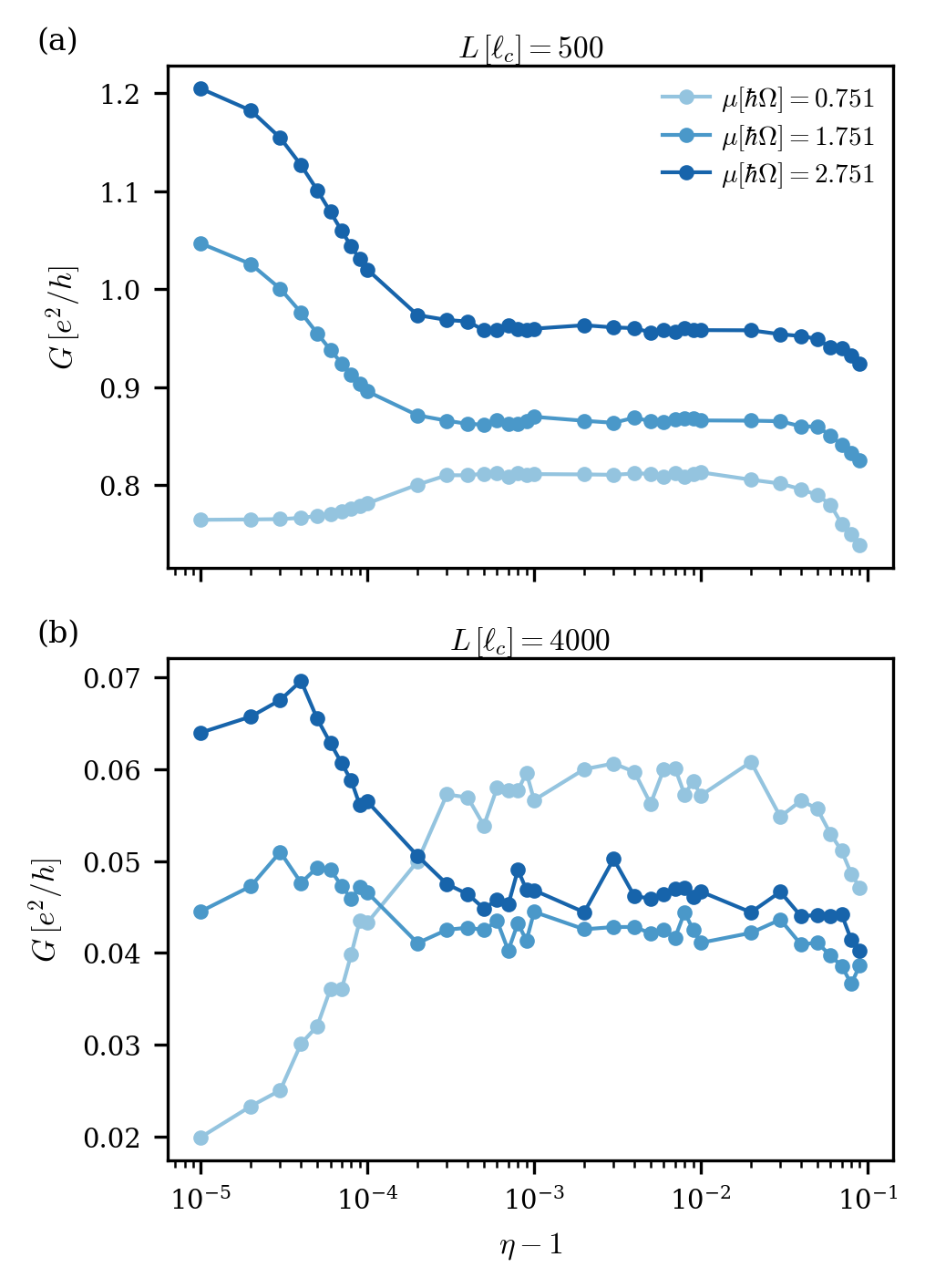}
\par\end{centering}
\caption{{\bfseries Conductance near integer flux.}
The conductance \(G\) is shown as a function of the dimensionless axial flux
\(\eta=\Phi/\Phi_0\), for flux values close to \(\eta_c=1\). The two panels
correspond to wire lengths (a) \(L/\ell_c=500\) and
(b) \(L/\ell_c=4000\), shown for chemical potentials
\(\mu/(\hbar\Omega)=0.751,\,1.751,\) and \(2.751\). The remaining parameters are the same as in Fig.~\ref{fig:conductance-L-eta}.}
\label{fig:WAL}
\end{figure}

We next focus on the conductance peaks that appear near integer values of
the dimensionless axial flux, shown in Fig.~\ref{fig:WAL}. These peaks are the remnant of
weak anti-localization (WAL) in the TINW surface states. In the usual
picture, WAL arises from the interference between pairs of time-reversed
scattering paths. For surface Dirac states, a closed scattering trajectory
around the Fermi surface acquires a Berry phase of \(\pi\), leading to
destructive interference between backscattered paths. This suppresses
localization and enhances the conductance near the time-reversal-invariant
flux values.

The strength of the WAL peak depends on the number of available conducting
channels. Close to the Dirac point, only a small number of angular-momentum
channels cross the chemical potential, so the system realizes a discrete
version of the usual WAL mechanism. As the chemical potential is increased,
more channels become available for scattering, and the WAL peak becomes slightly more
pronounced. This trend is seen in Fig.~\ref{fig:WAL}a, where the peak near
\(\eta=1\) grows as \(\mu\) is shifted away from the Dirac point. Remarkably, the lowest chemical potential shows
a distinct behavior: instead of a WAL peak, the conductance develops a weak
localization dip. This behavior may be attributed to the small number of
available angular-momentum channels, that are unable to sustain the adiabatic Berry phase.

For longer wires, the WAL peak is suppressed as the system crosses over to
the strongly localized regime. As this suppression takes place, the peak also
splits, as shown in Fig.~\ref{fig:WAL}b, while the lowest chemical potential retains the weak-localization dip.

The dependence on chemical potential can also be seen directly in the
two-dimensional conductance map of Fig.~\ref{fig:conductance-mu-eta}c.
There, the WAL peak appears near \(\eta=1\), but its strength is not uniform
within a conductance diamond. Moving away from the Dirac point, the number of
available angular-momentum channels increases, and the WAL peak generally
becomes stronger. However, this trend is interrupted near the edges of the
diamonds. At values \(\tilde{\mu}\sim m+1/2\), with \(m\in\mathbb{Z}\), the
chemical potential lies near a van-Hove singularity, where the density of
states is strongly enhanced. Near these points the WAL peak is suppressed.
Thus, within each diamond, the WAL feature reflects a competition between
the increasing number of available scattering channels and the suppression
near the van-Hove singularities.


\subsection{Critical exponent of the helical mode}

We now quantify the delocalization transition associated with the helical
mode at half flux quantum. As shown in
Figs.~\ref{fig:conductance-L-eta} and
\ref{fig:conductance-mu-eta}, the system remains conducting at
\(\eta\in\mathbb{Z}+1/2\), where the dimensionless conductance
\(\tilde{G}_L=G_L/(e^2/h)\) approaches unity in the long-wire limit. Away
from this point, the conductance eventually decays with increasing system
length, indicating localization.

To extract the localization length, we focus on the asymptotic localized
regime. We identify a length scale \(L_0\) beyond which the conductance drops
below the conductance quantum and follows an exponential decay,
\[
G_L = G_{L_0}\exp\left[-(L-L_0)/\xi\right],
\]
where \(\xi\) is the localization length
\cite{pichard1981finite1,pichard1981finite2,mackinnon1983scaling}. 

\begin{figure*}[t]
\begin{centering}
\includegraphics[width=0.95\textwidth]{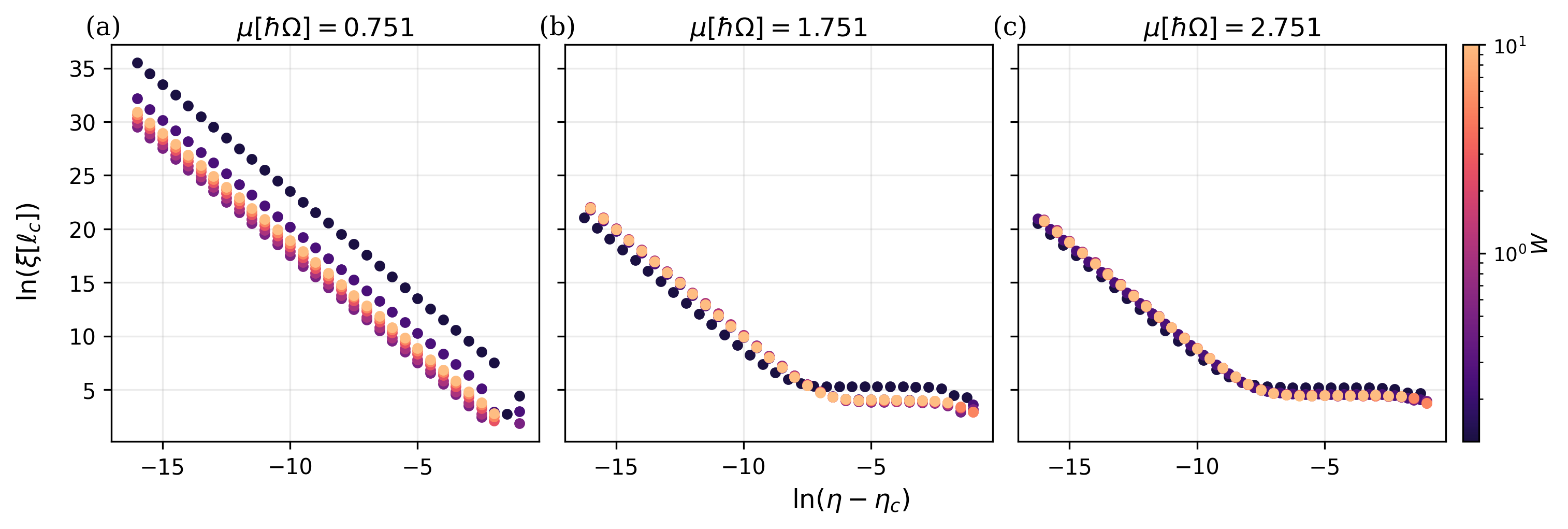}
\par\end{centering}
\caption{{\bfseries Delocalization transition near half flux quantum.}
The localization length \(\xi\), measured in units of \(\ell_c\), is extracted
from the exponential decay of the conductance and plotted as
\(\ln(\xi/\ell_c)\) versus \(\ln(\eta-\eta_c)\), with
\(\eta_c=1/2\). The three panels correspond to
(a) \(\mu=0.751\,\hbar\Omega\),
(b) \(\mu=1.751\,\hbar\Omega\), and
(c) \(\mu=2.751\,\hbar\Omega\). Different colors denote different disorder
strengths \(W\). The fitted critical exponent is
\(\nu=2.00\pm0.01\) for all values of \(\mu\) and \(W\) shown.}
\label{fig:critical_exponent}
\end{figure*}

At the critical point, \(\eta_c=1/2\), the protected helical state remains
extended, and therefore \(\xi(\eta_c)\) diverges. For nearby flux values, the
localization length follows
\[
\xi(\eta)\sim |\eta-\eta_c|^{-\nu},
\]
with \(\eta_c=1/2\). The extracted localization lengths are shown in
Fig.~\ref{fig:critical_exponent}. For all chemical potentials and disorder
strengths shown, the data are consistent with the same critical exponent,
\[
\nu = 2.00 \pm 0.01 .
\]

We find that this exponent is insensitive to the disorder strength over a
wide range of values. The collapse of the extracted slopes indicates that the
divergence of \(\xi\) is controlled by the distance from half flux quantum,
rather than by the microscopic strength of the disorder.

At low chemical potential, where only a single conducting channel is
available, the system is effectively one-dimensional. Away from
\(\eta=1/2\), the deviation from half flux opens a gap in the helical
spectrum, allowing backscattering and localization. The exponent
\(\nu=2\) saturates the Harris criterion in one dimension. At
higher chemical potentials, additional channels are present at short
lengths, but disorder localizes these non-protected modes. The long-distance
transport is then again governed by the helical channel, leading to the same
critical exponent.


\section{Methods}

We consider the surface Hamiltonian of the TINW,
\[
\hat{H}_{\text{nw}}+\hat{V},
\qquad
\hat{H}_{\text{nw}}
=
\hbar\Omega
\left[
\sigma_z\left(\hat{\ell}+\frac{1}{2}-\eta\right)
-\sigma_y \hat{k}R
\right],
\]
with \(\hat{V}=V(z,\varphi)\sigma_0\) the disorder potential. Here
\(\sigma_j\), \(j=x,y,z\), are Pauli matrices in the spin basis,
\(\sigma_0\) is the \(2\times2\) identity matrix,
\(\hat{k}=-i\partial_z\), and \(\hat{\ell}=-i\partial_\varphi\). The
dimensionless axial flux is \(\eta=\Phi/\Phi_0\), with \(\Phi_0=h/e\).~\cite{haver2023electromagnetic}.

The disorder is modeled as a sequence of impurities located at positions
\(z_s\), \(s=1,\ldots,N\), with a separable potential
\[
V(z,\varphi)
=
\sum_s V_s Z(z-z_s)Y(\varphi).
\]
The angular dependence is expanded as
\[
Y(\varphi)=2\sum_{m\in\mathbb{Z}}Y_m e^{im\varphi},
\qquad
Y_{-m}=Y_m^* .
\]
The impurity strengths \(V_s\) are drawn independently from a uniform
distribution in the interval \([-W,W]\). The distance between consecutive
impurities is also random: \(\Delta z_s=z_s-z_{s-1}\) is drawn independently
from a uniform distribution on \([0,2\ell_c]\), so that the mean impurity
spacing is \(\ell_c\). We also include an independent random angular shift
\(\Delta\varphi_s\), uniformly distributed over \([-2\pi,2\pi]\), between
successive scattering events.

We compute the conductance using a scattering-matrix approach. The
quasi-one-dimensional problem is reduced to a disordered Kronig-Penney-type
model~\cite{sanchez1994suppression}, in which electrons propagate freely
between impurities and scatter at each impurity. This formulation allows us
to access transport over large system lengths without multiplying unstable
transfer matrices directly.

For a single scatterer at \(z=z_s\), the scattering state at energy
\(\mu>0\) is written on the two sides of the impurity as
\begin{align}
\nonumber
|\Psi_\mu(z<z_s,\varphi)\rangle
&=
\sum_{\ell,\alpha}
a_{\ell}^{\alpha}
e^{i\alpha k_\ell z}
e^{i\ell\varphi}
|\Psi_{\ell\alpha}\rangle,\\
|\Psi_\mu(z>z_s,\varphi)\rangle
&=
\sum_{\ell,\alpha}
b_{\ell}^{\alpha}
e^{i\alpha k_\ell z}
e^{i\ell\varphi}
|\Psi_{\ell\alpha}\rangle .
\label{eq:states}
\end{align}
Here \(a_\ell^\alpha\) and \(b_\ell^\alpha\) are the amplitudes of the
orbital-angular-momentum channel \(\ell\in\mathbb{Z}\), with longitudinal
momentum \(\alpha k_\ell\), where \(k_\ell>0\) and \(\alpha=\pm\) labels
right- and left-moving modes. The momentum \(k_\ell\) lies on the Fermi
surface and is determined by
\[
\tilde{\mu}
=
\sqrt{
\left(\ell+\frac{1}{2}-\eta\right)^2
+
\tilde{k}_\ell^2
}.
\]
The spinors are written as
\[
|\Psi_{\ell\alpha}\rangle
=
U_{\ell\alpha}^{\dagger}|+\rangle,
\qquad
U_{\ell\alpha}
=
\frac{1}{\sqrt{k_\ell}}
\exp\left(
\frac{i}{2}\sigma_x\theta_{\ell\alpha}
\right),
\]
with
\[
\theta_{\ell\alpha}
=
\arctan\left(
\ell+\frac{1}{2}-\eta,
\alpha\tilde{k}_\ell
\right),
\]
where \(\arctan(x,y)\) is the four-quadrant inverse tangent and
\(|+\rangle=(1,0)^T\). The states are normalized by the group velocity
\(v(k_\ell)\propto k_\ell\), so that each mode carries the same probability
flux, as required for transport calculations~\cite{bruus2004many}.

For a single impurity,
\[
V(z,\varphi)=V_s Z(z)Y(\varphi),
\qquad
Z(z)=\hbar\Omega R\delta(z),
\]
integrating the Dirac equation across the impurity gives
\begin{align}
\nonumber
&
i\sigma_y
\left(
|\Psi(z_s^+,\varphi)\rangle
-
|\Psi(z_s^-,\varphi)\rangle
\right)
\\
&
+
V_s
\sum_m Y_m e^{im\varphi}
\left(
|\Psi(z_s^+,\varphi)\rangle
+
|\Psi(z_s^-,\varphi)\rangle
\right)
=0 .
\label{eq:continuity}
\end{align}
Substituting Eq.~\eqref{eq:states}, we obtain
\begin{align}
\nonumber
\sum_{\ell\alpha}
\bigg[
&
i\sigma_y
\left(
b_\ell^\alpha-a_\ell^\alpha
\right)
\\
&
+
V_s
\sum_m
Y_m e^{im\varphi}
\left(
b_\ell^\alpha+a_\ell^\alpha
\right)
\bigg]
e^{i\ell\varphi}
|\Psi_{\ell\alpha}\rangle
=0 .
\end{align}
Projecting onto the basis elements by multiplying from the left by
\(\langle\Psi_{\ell'\alpha'}|e^{-i\ell'\varphi}\) and averaging over the
angle, \((2\pi)^{-1}\int_0^{2\pi}d\varphi\), gives
\begin{align}
\nonumber
&
\sum_{\ell\alpha}
\left(
\mathcal{Y}_{\ell'\alpha';\ell\alpha}\delta_{\ell',\ell}
+
V_s
\sum_m
\mathcal{V}_{\ell'\alpha';\ell\alpha}^{(m)}
\delta_{\ell'-m,\ell}
\right)
b_\ell^\alpha
\\
&=
\sum_{\ell\alpha}
\left(
\mathcal{Y}_{\ell'\alpha';\ell\alpha}\delta_{\ell',\ell}
-
V_s
\sum_m
\mathcal{V}_{\ell'\alpha';\ell\alpha}^{(m)}
\delta_{\ell'-m,\ell}
\right)
a_\ell^\alpha ,
\label{eq:matching}
\end{align}
where
\begin{align}
\nonumber
\mathcal{V}_{\ell'\alpha';\ell\alpha}^{(m)}
&=
Y_m
\langle\Psi_{\ell'\alpha'}|\Psi_{\ell\alpha}\rangle
=
\frac{Y_m}{\sqrt{k_{\ell'}k_\ell}}
\cos\left(
\frac{\theta_{\ell'\alpha'}-\theta_{\ell\alpha}}{2}
\right),
\\
\nonumber
\mathcal{Y}_{\ell'\alpha';\ell\alpha}
&=
i
\langle\Psi_{\ell'\alpha'}|
\sigma_y
|\Psi_{\ell\alpha}\rangle
=
\frac{-i}{\sqrt{k_{\ell'}k_\ell}}
\sin\left(
\frac{\theta_{\ell'\alpha'}+\theta_{\ell\alpha}}{2}
\right).
\end{align}
where, for short-range impurities, $Y_m=1$ for all $m$. These matrix elements satisfy
\[
\mathcal{V}_{\ell'\alpha';\ell\alpha}^{(m)*}
=
\mathcal{V}_{\ell\alpha;\ell'\alpha'}^{(-m)},
\qquad
\mathcal{Y}_{\ell'\alpha';\ell\alpha}^{*}
=
-\mathcal{Y}_{\ell\alpha;\ell'\alpha'} .
\]
For pairs of channels satisfying
\[
\ell+\frac{1}{2}-\eta
=
-\left(\ell'+\frac{1}{2}-\eta\right),
\]
the backscattering matrix element vanishes,
\[
\mathcal{V}_{\ell'\alpha';\ell,-\alpha}^{(m)}=0 .
\]
In addition, \(\mathcal{Y}_{\ell\alpha;\ell,-\alpha}=0\).

Equation~\eqref{eq:matching} determines the transfer matrix \(M_s\) of a
single impurity,
\[
\boldsymbol{b}=M_s\boldsymbol{a},
\]
where
\[
\boldsymbol{a}=(a^+,a^-)^T,
\qquad
\boldsymbol{b}=(b^+,b^-)^T .
\]
Here \(a^\alpha\) and \(b^\alpha\) are row vectors whose elements are the
amplitudes \(a_\ell^\alpha\) and \(b_\ell^\alpha\) over the allowed
\(\ell\)-channels. Equivalently, one may define the scattering matrix
\[
S_s=
\begin{pmatrix}
r_s & t_s'\\
t_s & r_s'
\end{pmatrix},
\qquad
\boldsymbol{c}^{\text{out}}=S_s\boldsymbol{c}^{\text{in}},
\]
where
\[
\boldsymbol{c}^{\text{in}}=(a^+,b^-)^T,
\qquad
\boldsymbol{c}^{\text{out}}=(a^-,b^+)^T .
\]
The submatrices \(r_s,r_s'\) describe reflection from the left and from the
right, while \(t_s,t_s'\) describe transmission from left to right and from
right to left. The transfer matrix can be written in terms of these
submatrices as
\[
M_s=
\begin{pmatrix}
(t_s^\dagger)^{-1} & r_s' t_s'^{-1}\\
-t_s'^{-1}r_s & t_s'^{-1}
\end{pmatrix}.
\]

For multiple impurities, \(s=1,\ldots,N\), the electron propagates freely
between consecutive scatterers. Formally, the total transfer matrix is
\[
\mathcal{M}_N=\prod_{s=N}^{1}M_s\tilde{M}_s ,
\]
where \(M_s\) is the transfer matrix of impurity \(s\), and \(\tilde{M}_s\)
describes clean propagation over the distance
\(\Delta z_s=z_s-z_{s-1}\) and angular shift \(\Delta\varphi_s\) between
successive scatterers. The propagation matrix is obtained by setting
\(r_s,r_s'\to0\) and replacing
\[
t_s,t_s'
\to
\exp\left(
i\mathcal{K}\Delta z_s
\pm
i\mathcal{L}\Delta\varphi_s
\right),
\]
with
\[
\mathcal{K}
=
\operatorname{diag}
(k_{\ell_{\text{min}}},\ldots,k_{\ell_{\text{max}}}),
\qquad
\mathcal{L}
=
\operatorname{diag}
(\ell_{\text{min}},\ldots,\ell_{\text{max}}).
\]
Because products of transfer matrices contain exponentially large and small
eigenvalues, this formulation is numerically unstable
\cite{tamura1991conductance,Bardarson2018}. We therefore work directly with
the scattering matrices and combine them using the composition rule
\[
\mathcal{S}_N
=
\bigotimes_{s=N}^{1}
S_s\otimes\tilde{S}_s ,
\]
where \(\tilde{S}_s\) is the scattering matrix for clean propagation between
scatterers. For two scatterers,
\begin{align}
&S_2\otimes S_1 \equiv
\\
&\begin{pmatrix}
r_1+t_1'r_2(1-r_1'r_2)^{-1}t_1
&
t_1'(1-r_2r_1')^{-1}t_2'
\\
t_2(1-r_1'r_2)^{-1}t_1
&
r_2'+t_2r_1'(1-r_2r_1')^{-1}t_2'
\end{pmatrix}.
\nonumber
\end{align}
The conductance is then obtained from the Landauer formula,
\[
G=\frac{e^2}{h}\operatorname{Tr}(t^\dagger t),
\]
where \(t\) is the total transmission matrix extracted from
\(\mathcal{S}_N\).

Finally, we include the leading evanescent states in the scattering
calculation. These modes have imaginary momentum,
\(k_\ell=i\kappa_\ell\), and decay as
\(\Psi_\mu(z,\varphi)\sim e^{-\kappa_\ell z}\). Since they do not carry
current, they are excluded from the final transmission and reflection
matrices entering the Landauer formula. Nevertheless, they modify the
scattering amplitudes of the propagating modes, especially when the chemical
potential lies close to the bottom of an excited subband
\cite{cahay1990influence,bandyopadhyay1991role,tamura1991conductance}.

\section{Conclusions} 

We have shown that disordered topological insulator nanowires provide a
controlled setting for studying flux-driven localization and delocalization
in the presence of symmetry-protected transport. By following the conductance
as a function of wire length, magnetic flux, and chemical potential, we
resolved the crossover from ballistic and interference-dominated transport
to the asymptotic localized regime. This allowed us to extract the
localization length \(\xi\) and determine its scaling with the dimensionless
flux \(\eta\).

Our central result is the characterization of a flux-driven delocalization
transition near half flux quantum. At this point, the protected helical mode
remains transmitting, while away from it the localization length diverges as
\(\xi\sim|\eta-1/2|^{-\nu}\), with a robust critical exponent
\(\nu\simeq 2\). This exponent is independent of the chemical potential and
disorder strength considered here, and differs from the critical exponent of
the integer quantum Hall transition, suggesting that the flux-driven
transition in TINWs may belong to a different universality class. In
addition, near integer flux quanta, the conductance feature evolves from a
weak-localization dip at low chemical potential to a weak anti-localization
peak at higher chemical potential, indicating a channel-number-dependent
crossover between distinct interference regimes.

Other sources of disorder in TINWs include magnetic impurities, which break
time-reversal symmetry and can localize the helical surface states, thereby
suppressing the characteristic anti-localization response. In addition, local
curvature or geometric inhomogeneities \cite{Bardarson2018} can generate
quasi-bound states that further enhance localization tendencies
\cite{saxena2022electronic}. These effects highlight possible directions for
future studies of disorder-driven localization in topological nanowires.
Beyond their fundamental relevance, our results may provide a useful
reference point for studying TINWs in gatemon-type superconducting devices
\cite{yavilberg2019differentiating,schmitt2022integration}, where the
interplay between disorder, topology, and superconducting proximity is
important.


\section{Acknowledgments}
We are grateful to D.~Dahan for useful discussions. We would like to thank the Israeli Ministry of Innovation, Science and Technology for support under the UK/Israel Research Collaboration Grant QuantumGate.

\bibliography{bibfile} 

\end{document}